\documentclass[preprint,showpacs,preprintnumbers,amsmath,amssymb]{revtex4}
\usepackage{graphicx}
\usepackage{dcolumn}
\usepackage{bm}

\begin{document}

\preprint{APS/123-QED}
\title{
Determination of  the parameters of a Skyrme type
effective interaction using the simulated annealing approach }

\author{B. K. Agrawal,  S. Shlomo,  and V. Kim Au}
\affiliation{Cyclotron Institute, Texas A\&M University,
College Station, TX 77843, USA}

\begin{abstract}
We implement for the first time  the simulated annealing method (SAM)
to the problem of searching for  the global minimum in the hyper-surface
of the chi-square function which depends on the values of the parameters
of a Skyrme type effective nucleon-nucleon  interaction.  We undertake
a realistic case of fitting the values of the  Skyrme parameters to an
extensive set of experimental data on the ground state properties of
many  nuclei ranging from normal to exotic ones.  The set of experimental
data used in our fitting procedure includes the radii for the valence
$1d_{5/2}$ and $1f_{7/2}$ neutron orbits in the $^{17}$O and $^{41}$Ca
nuclei, respectively,  and the breathing mode energies for several nuclei,
in addition to the typically used data on binding energy, charge radii
and spin-orbit splitting.  We also include in the fit the critical density
$\rho_{cr}$ and  further constrain the values of the Skyrme parameters by
requiring that (i) the quantity $P = 3\rho \frac{dS}{d\rho}$, directly
related to the slope of the symmetry energy $S$, must be positive for
densities up to $3\rho_0$ (ii) the enhancement factor $\kappa$, associated
with the isovector giant dipole resonance,  should lie in the range of
$0.1 - 0.5$ and (iii) the Landau parameter $G_0^\prime$ is positive at
$\rho = \rho_0$.  We provide  simple but consistent schemes to account
for the center of mass corrections to the binding energy and charge radii.

\end{abstract}  \maketitle

\section{Introduction}
\label{intro}
Since the pioneering work of Brink and Vautherin \cite{Vautherin72},
continuous efforts have been made to readjust the parameters of the
Skyrme type effective nucleon-nucleon interaction to better reproduce
experimental data.  Most of the parameters of the Skyrme interactions
available in the literature were obtained by fitting the Hartree-Fock (HF)
results to experimental data on bulk properties of a few stable closed
shell nuclei.  Only recently, several families  of the Skyrme parameters
e.g., SkI1-6, SLy1-10, SKX and SkO \cite{Reinhard95,Chabanat97,Chabanat98,
Brown98, Dean99, Bender03} were obtained by fitting HF results to the
experimental data on the bulk properties of nuclei ranging from the
$\beta-$stable nuclei to those near the proton and/or neutron drip
lines. In the SKX interaction, to yield appropriately the values for
the binding energy differences between mirror nuclei also referred to 
as the Coulomb displacement energy (CDE), the contribution of the
Coulomb exchange term in the HF equations is ignored and the direct
Coulomb term is evaluated by replacing the point proton distribution
by its charge distribution.  However, it has been further shown in
Ref. \cite{Stone03} that the SKX interaction is not suitable for studying
the properties of neutron stars.  Since, for the SKX interaction,
the quantity 
\begin{equation} P = 3\rho \frac{dS}{d\rho}, \label{P}
\end{equation} 
which is directly related to the slope of the symmetry
energy coefficient $S$, becomes negative for nuclear matter densities
$\rho$ well below $3\rho_0$ ($\rho_0 = 0.16$ fm$^{-3}$ is the saturation density).  On the other
hand, the SkI1-6, SLy1-10 and SkO Skyrme interactions  are found to
be suitable  for the study of neutron stars \cite{Stone03}. But, these
families of Skyrme interactions significantly underestimate the values of
the CDE for mirror nuclei, since the Coulomb exchange term was included.
Thus, it is desirable to have a unified interaction which includes
the merits of several families of the Skyrme interactions as mentioned
above. One can further  enhance the applicability of the Skyrme type
effective nucleon-nucleon interaction by imposing  certain constraints
as discussed below.

The aim of this work is twofold. We implement, for the first time,
the simulated annealing method  (SAM) \cite{Patrik83,NR92} to fit the
values of the  Skyrme parameters  and develop a more realistic Skyrme
type effective interaction.  The SAM is an elegant technique  for 
optimization problems of large scale, in particular, where a desired
global extremum is hidden among many local extrema. 
This method has been found
to be an extremely useful tool for a wide variety of minimization
problems of large non-linear systems in many different areas of science
(e.g., see Refs. \cite{Patrik84,Ingber89,Cohen94}).  Very recently
\cite{Burvenich02,Burvenich04}, the SAM was used to generate some initial
trial parameter sets for the  point coupling variant of the relativistic
mean field model.  In the present context, we use the SAM to determine the
values of the Skyrme parameters by searching for the global minimum  in the
hyper-surface of the  $\chi^2$ function given as,
\begin{equation}
\chi^2 =  \frac{1}{N_d - N_p}\sum_{i=1}^{N_d} \left (\frac{ M_i^{exp} -
M_i^{th}}{\sigma_i}\right )^2 \label {chi2}
\end{equation}
where, $N_d$ and $N_p$ are the number of  experimental data points
and the number of fitted  parameters, respectively, $\sigma_i$ is the 
theoretical error and $M_i^{exp}$ and $M_i^{th}$ are the
experimental and the corresponding theoretical values, respectively,
for a given observable.  The values of $\chi^2 $ depends on the Skyrme
parameters, since, the $M_i^{th}$ in Eq. (\ref{chi2})  is calculated using
the HF approach with a  Skyrme type effective nucleon-nucleon  interaction.

Toward the purpose of obtaining a more realistic parameterization of the 
Skyrme interaction, we apply the SAM to fit the HF results  to an
extensive set of experimental data for the binding energy, charge
radii, spin-orbit splitting and root mean square (rms) radii of valence
neutron orbits. Our data set used in the fit consists of 14
spherical nuclei, namely, $^{16}$O, $^{24}$O, $^{34}$Si, $^{40}$Ca,
$^{48}$Ca, $^{48}$Ni, $^{56}$Ni, $^{68}$Ni, $^{78}$Ni, $^{88}$Sr,
$^{90}$Zr, $^{100}$Sn, $^{132}$Sn and $^{208}$Pb. We also include in
our fit the experimental data for the breathing mode energies for the
$^{90}$Zr, $^{116}$Sn, $^{144}$Sm and $^{208}$Pb nuclei.  In addition,
we  include in the fit,  the critical density $\rho_{cr}$ which is
determined from   the stability conditions for the  Landau parameters.
\cite{Margueron02, Agrawal04}.  We further constrain the values of the
Skyrme parameters by requiring that (i) the quantity $P$ must be positive
for densities up to $ 3\rho_0$; a condition imposed by neutron star models
\cite{Stone03},  (ii) the enhancement factor $\kappa$, associated with
the Thomas-Reiche-Kuhn (TRK) sum rule for the isovector giant dipole
resonance, lies  in the range of $0.1 - 0.5$ \cite{Chabanat97,Berman75,
Krivine80, Keh91} and (iii) the Landau parameter $G_0^\prime$, crucial
for the spin properties of finite nuclei and nuclear matter, should be
positive at $\rho = \rho_0$ \cite{Keh91,Keh-fei91}.  We also provide
simple but  consistent schemes to account appropriately for the CDE and
the center of mass corrections to the binding energy and charge radii.
In order to check the reliability of the  proposed Skyrme interactions for
the study of high density matter, we have examined in detail the behavior
of the symmetry energy  and the nature of the equation of state (EOS)
for pure neutron matter at densities relevant for the neutron star models.

We have organized our paper as follows. In Sec. \ref{skedf} we briefly
outline the form of the Skyrme nucleon-nucleon (NN) effective interaction
and the corresponding energy density functional adopted in the present
work. In this section, we also  provide a viable strategies for the
calculations of CDE  and the center of mass corrections to the total
binding energy and charge radii.  In Sec. \ref{sknm} we provide the
relations between the Skyrme parameters and the various nuclear matter
properties, which we have used to implement the SAM algorithm as described
in Sec. \ref{sa}.  The set of the experimental data along with the
theoretical errors and the constraints  used in the  fit to determine
the values of the Skyrme parameters are given in Sec. \ref{bain}.
In Sec. \ref{res} we present our results for the two different  fits carried out
in this work.  Finally, in Sec. \ref{conc} we summarize our main results
and discuss the scope for the further improvement of the present work.

\section{Skyrme energy density functional}
\label{skedf}
In this work we adopt  the following form for the Skyrme type effective nucleon-nucleon
interaction \cite{Vautherin72,Chabanat97},
\begin{eqnarray}
\label{v12}
&V_{12}&= t_0\left (1+x_0 P_{12}^\sigma\right )\delta({\bf r}_1-{\bf
r}_2)\nonumber\\
&&+\frac{1}{2}t_1\left (1+x_1 P_{12}^\sigma\right ) 
\times \left[\overleftarrow{k}_{12}^2\delta({\bf r}_1-{\bf r}_2)+\delta({\bf
r}_1-{\bf r}_2)\overrightarrow{k}_{12}^2\right] \nonumber\\
&&+t_2\left (1+x_2 P_{12}^\sigma\right )\overleftarrow{k}_{12}\delta({\bf r}_1-{\bf r}_2)\overrightarrow{k}_{12} \nonumber\\
&&+ \frac{1}{6}t_3\left (1+x_3 P_{12}^\sigma\right )\rho^\alpha\left(\frac{{\bf r}_1+{\bf r}_2}{2}\right )
\delta({\bf r}_1-{\bf r}_2) \nonumber\\
&&+iW_0\overleftarrow{k}_{12} \delta({\bf r}_1-{\bf
r}_2)(\overrightarrow{\sigma_1}+
\overrightarrow{\sigma_2})\times \overrightarrow{k}_{12}
\end{eqnarray}
where, $t_i$, $x_i$, $\alpha$ and $W_0$ are the parameters of the 
interaction and  $P_{12}^\sigma$ is the spin exchange
operator, $\overrightarrow{\sigma}_i$ is the
Pauli spin operator, $\overrightarrow{k}_{12} = -i(
\overrightarrow{\nabla}_1-\overrightarrow{\nabla}_2)/2$
and $\overleftarrow{k}_{12} =
-i(\overleftarrow{\nabla}_1-\overleftarrow{\nabla}_2)/2\, .$ Here,
the   right and left arrows indicate that the momentum operators act
on the right and  on the left,  respectively.  The corresponding mean-field
$V_{HF}$ and the total energy $E$ of the system are given by, 
\begin{equation} V_{HF}=\frac{\delta \cal H}{\delta \rho}, \qquad
 E=\int {\cal H}(r)d^3r \end{equation}
where, the Skyrme energy density functional ${\cal H}(r)$, obtained using 
Eq. (\ref{v12}),  is given by \cite{Vautherin72, Chabanat97},
\begin{equation}
\label{Hden}
{\cal H} = {\cal K} + {\cal H}_0  +{\cal H}_3+{\cal H}_{\rm eff} +{\cal H}_{\rm fin}
+ {\cal H}_{\rm so} +{\cal H}_{\rm sg}+{\cal H}_{\rm Coul}
\end{equation}
where, ${\cal K} = \frac{\hbar^2}{2m}\tau$ is the kinetic energy term,
${\cal H}_0$ is the zero-range  term, ${\cal H}_3$ the density dependent
term, ${\cal H}_{\rm eff}$ an effective-mass term, ${\cal H}_{\rm fin}$
a finite-range term, ${\cal H}_{\rm so}$ a spin-orbit term, ${\cal
H}_{\rm sg}$ a term due to tensor coupling  with spin and gradient and
${\cal H}_{\rm Coul}$ is the contribution to the energy density due to the
Coulomb interaction. For the Skyrme interaction of Eq. (\ref{v12}), we have,

\begin{equation}
{\cal H}_0 = \frac{1}{4}t_0\left [(2 + x_0)\rho^2 - (2x_0 + 1)(\rho_p^2 + \rho_n^2)\right ],
\end{equation}
\begin{equation}
{\cal H}_3 = \frac{1}{24}t_3\rho^\alpha\left [(2 + x_3)\rho^2 - (2x_3 + 1)(\rho_p^2 + \rho_n^2)\right ],
\end{equation}
\begin{equation}
{\cal H}_{\rm eff} = \frac{1}{8} \left [t_1 (2 + x_1) + t_2 (2 + x_2)\right ]
\tau\rho
+ \frac{1}{8}\left[ t_2(2x_2+1) -t_1(2x_1+1)\right] (\tau_p\rho_p + \tau_n\rho_n), 
\end{equation}
\begin{eqnarray}
{\cal H}_{\rm fin}&=&\frac{1}{32}\left [3t_1 (2 + x_1) - t_2 (2 + x_2)\right ](\nabla\rho)^2 \nonumber\\
&& - \frac{1}{32}\left[ 3t_1(2x_1+1) +  t_2(2x_2+1) \right] \left [ (\nabla\rho_p)^2 + (\nabla\rho_n)^2\right], \\
{\cal H}_{\rm so}&=&\frac{W_0}{2}\left [ {\bf J}\cdot\nabla\rho +
x_w({\bf J_p}\cdot{\bf \nabla}\rho_p
+{\bf J_n}\cdot{\bf \nabla}\rho_n)\right ],\\
{\cal H}_{\rm sg} &=&-\frac{1}{16}(t_1 x_1 + t_2 x_2){\bf J}^2 + 
\frac{1}{16}(t_1 - t_2) \left [{\bf J_p}^2 + {\bf J_n}^2 \right]. 
\label{Hsg}
\end{eqnarray}
Here, $ \rho = \rho_p + \rho_n,$ $\tau = \tau_p + \tau_n,$ {\rm and}
${\bf J}= {\bf J}_p + {\bf J}_n$ are the particle number density,
kinetic energy density and spin density with $p$ and $n$ denoting the
protons and neutrons, respectively.  We have used the value of $\hbar^2/2m = 20.734$
MeVfm$^2$ in our calculations. We would like to emphasize that we
have  included the contributions from the spin-density term as given by
Eq. (\ref{Hsg}) which is ignored in many Skyrme HF calculations. Although
the contributions from the Eq.  (\ref{Hsg}) to the binding energy and
charge radii are not very significant, they are very crucial for
the calculation of the Landau parameter $G_0^\prime$ \cite{Bender02}.

\subsection{Coulomb energy}
The contribution to the energy density (Eq. (\ref{Hden})) from the 
Coulomb interaction can be written as a sum of a direct and an exchange
terms,
\begin{equation}
{\cal H}_{\rm Coul}(r)={\cal H}_{\rm Coul}^{dir}(r)  +{\cal H}_{\rm
Coul}^{ex}(r).
\label{Hcoul}
\end{equation}
For the direct term it is common to adopt the expression
\begin{equation}
{\cal H}_{\rm Coul}^{dir}(r) =\frac{1}{2}e^2\rho_p(r)
\int \frac{\rho_p( r^\prime)d^3r^\prime}{\mid \bf r - \bf r^\prime\mid },
\label{Hdir}
\end{equation}
and for the corresponding exchange term to use the Slater approximation 
\cite{Giannoni80},
\begin{equation}
{\cal H}_{\rm Coul}^{ex}(r) =-\frac{3}{4}e^2\rho_p(r) \left ( \frac{3 \rho_p(r)}{\pi}\right )^{1/3}.
\label{Hex}
\end{equation}
It is very important to emphasize that the definitions of Eqs. (\ref{Hdir})
and (\ref{Hex}) are not for the bona fide direct and exchange terms,
since each of them includes the contributions of the self-interaction,
which appear in opposite signs and cancel out in Eq. (\ref{Hcoul}). Note,
in particular, that the direct term of Eq. (\ref{Hdir}) is proportional
to $Z^2$ and not to $Z(Z-1)$, as it should be for a direct term, see
a detailed discussion in Ref. \cite{Shlomo78}.  We point out that for
the CDE of mirror nuclei the magnitude of the self interaction term is
CDE/(2$Z$), i.e., one has a spurious increase in the calculated CDE of about
6.3\% and 2.5\% for the A=17 and 41 systems of mirror nuclei, respectively.

We recall that within the mean-field approximation, adjusted to reproduce
the experimental values of the charge rms radii, the calculated CDE
of analog states (obtained using Eq. (\ref{Hcoul})) are smaller than the 
corresponding experimental values by about 7\%. It was first shown in 
Ref. \cite{Shlomo82} that this discrepancy, also known as the Nolen-Schiffer 
anomaly  \cite{Nolen69}, can be explained by taking into account the 
contributions due to long range correlations (LRC) and due to the charge 
symmetry breaking (CSB) in the NN interaction, see also Ref. \cite{Agrawal01}. 
We add that for the mirror nuclei with A=17 and A=41, the LRC and 
the CSB each accounts for about half of the discrepancy between theory and 
experiment. Also, the magnitude of the bona fide exchange Coulomb term is 
about the same as that due to LRC, but with opposite sign. Therefore, 
neglecting the bona fide Coulomb exchange term does not resolved the 
discrepancy between theory and experiment, but can account for the 
contribution of LRC.  We add that in Ref. \cite{Brown00}, it was shown
that by ignoring the Coulomb exchange term in the form of  of Eq. (14)
in Eq. (12), i.e., by including only the Coulomb direct term in the form
of Eq. (13) (as is the case for the SKX interaction), one reproduces the
experimental values of the CDE. It should be clear that this is due to
the fact that by adopting the form of Eq. (13) for the Coulomb direct term
one not only neglects the bona fide Coulomb exchange term, but also adds
the spurious contribution of the self-interaction term. The unphysical
neglect of the bona fide Coulomb exchange term together with the spurious
contribution of the self-interaction term results in a contribution to
CDE which is similar in magnitude to that obtained from the LRC + CSB
terms. For simplicity, we will adopt in this work the form of Eq. (13)
for the Coulomb direct term.

\subsection { Center of mass corrections to the binding energy and charge radii}
\label{cmc}

The HF approach applied to finite nuclei violates the translational invariance,
introducing a spurious center of mass (CM) motion. Thus, one  must extract
the contributions of the center of mass motion to the binding energy B,
radii and other observables. An accurate way to restore the translational
invariance is to use the projection method.  But, it is numerically
very expensive. So, it is desirable to develop simple schemes for the
CM corrections to various observables.  Normally, one makes the CM
corrections only to the binding energy and not to the radii. However,
the CM corrections to the rms radii for light nuclei may be as large
as $2\%$ \cite{Shlomo78}.  In the present work we shall consider the CM
corrections to the binding energy as well as charge rms radii used to
fit the Skyrme parameters.

To account for the CM correction to the total binding energy, one must subtract
from it  the so-called CM energy given as,
\begin{equation}
E_{CM} = \frac{1}{2mA}\langle  \hat{\bf P}^2 \rangle
\label{ecm}
\end{equation}
where,  $\hat{\bf P} = - i \hbar \sum_{i=1}^A \nabla_i $ is the total
linear momentum operator.  Traditionally, one simplifies the computation
of Eq. (\ref{ecm}) by taking into account only the  one-body parts of it, which
can be easily achieved by replacing $\frac {1}{m}\rightarrow \frac{1}{m}
\left [ 1 - \frac {1}{A} \right ],$ in the kinetic energy term. In
this case, the effects of neglecting the two-body part of Eq. (\ref{ecm})
are compensated by renormalization of the force parameters. This may
induce in the forces an incorrect trend with respect to $A$ which
becomes visible in the nuclear matter properties. In fact, it is found
in Ref. \cite{Bender00} that an over simplified  treatment of $E_{CM}$
obtained  by renormalizing the nucleon mass appearing in the kinetic
energy term leads to a larger value of the surface energy coefficient
than those obtained using the full CM correction. This gives rise
to differences in the deformation energy which becomes quite pronounced
for the super deformed states. Very recently \cite{Agrawal04}, we also
find that a large value of the surface energy coefficient yield a smaller
value for the critical density. Thus, an appropriate and still simple
scheme to evaluate Eq. (\ref{ecm}) is highly desirable. We note, however,
that the  SLy6,
SLy7 and SLy10 interactions \cite{Chabanat98}  have been obtained by
evaluating Eq.  (\ref{ecm}) ( i.e., including the one and two-body CM terms
of  Eq. (\ref{ecm})).  In the harmonic oscillator (HO) approximation, $E_{CM}$
of Eq.  (\ref{ecm}) is given by,
\begin{equation}
E_{CM}^{osc} = \frac{3}{4}\hbar\omega 
\label{ecm-osc}
\end{equation}
A value of $\hbar\omega= 41 A^{-1/3}$  MeV is used
in many relativistic mean-field calculations
\cite{Lalazissis97,Ring97}.  An improved version for the  CM correction 
can be obtained by modifying the oscillator frequency as $\hbar\omega =
45A^{-1/3} - 25A^{-2/3}$ MeV, which has been used in Ref. \cite{Brown98} to
obtain the SKX interaction. Here, we employ  a simple but more consistent
scheme to evaluate the  $E_{CM}$ using the HO approximation.  We determine
the oscillator frequency $\hbar\omega$ appearing in Eq. (\ref{ecm-osc})
using the mean square  mass radii $\langle r^2\rangle $ calculated in the HF
approach  as,
\begin{equation}
\hbar\omega = \frac{\hbar^2}{mA\langle r^2\rangle }\sum_i[{\cal N}_i+
\frac{3}{2}],
\label{hw}
\end{equation}
where, the sum runs over all the occupied single-particle states for
the protons and neutrons and ${\cal N}_i$ is the oscillator quantum
number. We emphasize that this scheme is  quite reliable
even for the nuclei away from the $\beta$-stable line where the values of
the  rms radii deviate from the $A^{1/3}$ law.  We have calculated the total
binding energy for the SLy7 interaction using our simple scheme for
the CM correction, Eq. (\ref{hw}), and compare them with those given in
Ref. \cite{Chabanat98}, obtained by using one-body and two-body parts
of the Eq. (\ref{ecm}). For example, we find that for the  $^{16}$O,
$^{40}$Ca, $^{132}$Sn and $^{208}$Pb nuclei the total binding energy
$B = 128.65$ (128.55), 344.98 (344.90), 1102.38 (1102.77) and 1636.29
(1636.76) MeV, respectively, where the values in parenthesis are
taken from Ref. \cite{Chabanat98}. This clearly indicates that the CM
correction to the binding energy can be reliably estimated using Eq.
(\ref{hw}). We would also like to remark, however,  that the $E_{CM}$
calculated using the oscillator frequency as $\hbar\omega = 45A^{-1/3}
- 25A^{-2/3}$ MeV in Eqs. (\ref{ecm-osc}) overestimates the value of
binding energy in light nuclei (e.g., $^{16}$O and $^{40}$Ca) by about
$1 - 2$ MeV which is quite significant.

The mean-square  radius for the point proton distribution  corrected for the CM motion is obtained as \cite{Shlomo78},
\begin{equation}
\langle r^2_p\rangle = \langle r^2_p\rangle_{\rm HF}-\frac{3}{2\nu A},
\label{rpav}
\end{equation}
where, $\nu = m\omega/\hbar$ is the size parameter. Therefore,  the
corresponding mean-square charge radius to be fitted to the experimental
data is obtained as,
\begin{equation}
\langle r^2_{ch}\rangle = \langle r^2_p\rangle_{\rm HF}-\frac{3}{2\nu
A} +\langle r^2\rangle_p+ \frac{N}{Z}\langle r^2\rangle_n+\frac{1}{Z}
\left( \frac{\hbar}{mc}\right ) \sum_{nlj\tau}(2j+1)\mu_\tau \langle
{\bf \sigma}\cdot {l}\rangle_{lj},
\label{rc0}
\end{equation}
where, $\langle r^2\rangle_p$ and $\langle r^2\rangle_n$ are the
mean-squared radii of the proton and neutron charge distributions,
respectively.  The last term in Eq. (\ref{rc0}) is due to the  spin-orbit
effect \cite{Bertozzi72}. We use, $\langle r^2\rangle_n = -0.12$ fm$^2$
and  the recent \cite{Ingo03}  value of $\langle r^2\rangle_p =
0.801$ fm$^2$.

\subsection{Determination of the critical density}
\label{rhocr}
We use the  stability conditions of the Landau parameters for the
symmetric nuclear matter and pure neutron matter to calculate the
critical density $\rho_{cr}$ for the Skyrme type effective nucleon-nucleon
interactions.  The stability conditions  are given as \cite{Migdal67},
\begin{equation}
\label{ic}
{\cal A}_l > -(2l + 1),
\end{equation}
where, ${\cal A}_l$ stands for the Landau parameters $F_l$, $F_l^\prime$,
$G_l$ and $G_l^\prime$ for a given multipolarity $l$.  Skyrme interactions
only contain monopolar and dipolar contributions to the particle-hole
interaction so that all Landau parameters are zero for $l > 1$. Thus,
there are 12  different Landau parameters, i.e.,  $F_l$, $F_l^\prime$,
$G_l$ and $G_l^\prime$ ($l = 0, 1$) for the symmetric nuclear matter and
$F_l^{(n)}$, $G_l^{(n)}$ ($l= 0, 1$) for the pure neutron matter. Each
of these parameters must satisfy the inequality condition given by
Eq. (\ref{ic}). Explicit expressions for the Landau parameters in terms of
the Skyrme parameters can be found in Refs.  \cite{Margueron02,Krewald77}.
The critical density is nothing but the maximum density beyond which
at least one of the Landau Parameter does not satisfy  Eq. (\ref{ic}).
Following Ref. \cite{Margueron02}, one can obtain the  values of the
Landau parameters at any density for a given set of the Skyrme parameters.
Thus, for a given set of Skyrme parameters one can easily obtain the
value of $\rho_{cr}$.  As mentioned in Sec. \ref{intro}, 
we include $\rho_{cr}$ in the fit.

\subsection {Breathing mode energy}
We also include in our fit the experimental data on the breathing mode
energy for several nuclei.  We consider  the fully self-consistent values
for the breathing mode constrained  energy defined as,
\begin{equation}
E_{0} = \sqrt{\frac{m_{1}}{m_{-1}}}, 
\label{e0}
\end{equation} 
where $m_k$ are the energy moments 
\begin{equation}
\label{mk}
m_k=\int_{0}^{\infty}\omega^k  S(\omega) d\omega,
\end{equation}
of the strength function
\begin{equation}
S(\omega)=\sum_n\left |\langle n\mid F\mid 0\rangle\right |^2
\delta(\omega-\omega_n),
\end{equation}
for the monopole operator
$ F(r)=\sum_{i=1}^A f(r_i)$, with $f(r)=r^2$. 
The moments $m_k$ for $k= -1 $ and 1 appearing in  Eq. (\ref{e0}) can
be obtained  using the constrained HF (CHF) and the double commutator
sum rule, respectively \cite{Bohigas79,Colo04,Shlomo04}.  The moment  $m_1$
can be expressed in terms of the ground state density  $\rho$ as,
\begin{equation}
\label{m1} m_1=2\frac{\hbar^2}{m}\langle r^2\rangle,
\end{equation}
where, 
\begin{equation} \langle r^2\rangle = \int r^2\rho(r) d{\bf r}.
\end{equation} 
As described in detail in Ref. \cite{Bohigas79,Colo04,Shlomo04}, $m_{-1}$
can be evaluated via the CHF approach and is given as,
\begin{equation} \label{m-1} m_{-1}=\left
. \frac{1}{2}\frac{d}{d\lambda}\langle r^2_\lambda \rangle \right
|_{\lambda=0}
\end{equation} 
where, 
$ \langle r_\lambda^2\rangle = \langle \Phi_\lambda\left
| r^2\right |\Phi_\lambda \rangle, 
$ 
with $\Phi_\lambda$ being the HF solution to the constrained HF
Hamiltonian $ H - \lambda f$.

\section{Skyrme parameters and nuclear matter properties}
\label {sknm}
In this section we discuss the relationship between the
Skyrme parameters and the various quantities describing the  nuclear
matter. In the next section we use  these relations to 
implement the SAM algorithm.  The  Skyrme parameters $t_i$,
$x_i$ and $\alpha$ for a fixed value of $W_0$ can be expressed in terms
of the quantities associated with the symmetric nuclear matter as follows
\cite{Chabanat97,Margueron02,Gomez92}.
\begin{eqnarray}
\label{t0}
t_0&=&\frac{8}{\rho_{nm}}\left [\frac{\left (-{B/A}+\left (2m/m^*-3\right )\left (\hbar^2/10m\right )k_f^2\right )\left (\frac{1}{27}K_{nm}-\left (1-6m^*/5m\right )\left (\hbar^2/9m^*\right )
k_f^2\right )}{-{B/A}+\frac{1}{9}K_{nm}-\left (4m/3m^*-1\right )\left (\hbar^2/10m\right )k_f^2}\right .\\ \nonumber
&+&\left . \left (1-\frac{5m}{3m^*}\right )\frac{\hbar^2}{10m}k_f^2\right ],
\end{eqnarray}
\begin{equation}
t_1=\frac{2}{3}\left [T_0 + T_s\right ],
\end{equation}
\begin{equation}
t_2=t_1+\frac{8}{3}\left[\left (
\frac{1}{4}t_0+\frac{1}{24}t_3\rho_{nm}^\alpha\right )
\frac{2m^*}{\hbar^2}\frac{k_f}{\pi^2}+G_0^\prime\right ]
\frac{\hbar^2}{m^*\rho_{nm}},
\label{t2}
\end{equation}
\begin{equation}
t_3=\frac{16}{\rho_{nm}^{\alpha+1}}\frac{\left (-{B/A}+\left (2m/m^*-3\right )\left (\hbar^2/10m\right )k_f^2\right )^2}
{-{B/A}+\frac{1}{9}K_{nm}-\left (4m/3m^*-1\right )\left
(\hbar^2/10m\right )k_f^2}, \label{t3}
\end{equation}
\begin{equation}
x_0=\frac{4}{t_0\rho_{nm}}\left [\frac{\hbar^2}{6m}k_f^2 -\frac{1}{24}t_3
(x_3+\frac{1}{2})\rho_{nm}^{\alpha+1} + \frac{1}{24}\left (t_2\left
(4+5x_2\right )-3t_1x_1\right )\rho_{nm} k_f^2 -J \right]- \frac{1}{2},
\end{equation}
\begin{equation}
x_1=\frac{1}{t_1}\left [4\frac{\hbar^2 \kappa}{m\rho_{nm}}-t_2(2+x_2)\right ] -2,
\label{x1}
\end{equation}
\begin{equation}
x_2=\frac{1}{4t_2}\left [8T_0-3t_1-5t_2\right ],
\end{equation}
\begin{equation} 
x_3=-\frac{8}{\alpha t_3\rho_{nm}^{\alpha+1}}\left
[\frac{\hbar^2}{6m}k_f^2-\frac{1}{12}\left ((4+5 x_2)t_2-3t_1 x_1\right
)\rho_{nm} k_f^2-3 J +L\right ]-\frac{1}{2},
\label{x3}
\end{equation}
\begin{equation}
\alpha=\frac{{B/A}-\frac{1}{9}K_{nm}+\left (4m/3m^*-1\right )\left (\hbar^2/10m\right )k_f^2}
{-{B/A}+\left (2m/m^*-3\right )\left (\hbar^2/10m\right )k_f^2},
\label{alpha}
\end{equation}
where,

\begin{equation}
T_0=\frac{1}{8}\left (3t_1+(5+4 x_2)t_2\right )=
\frac{\hbar^2}{m\rho_{nm}}\left (\frac{m}{m^*}-1\right ),
\label{T0}
\end{equation}
\begin{equation}
T_s=\frac{1}{8}\left [9t_1-(5+4 x_2)t_2\right ],
\label{TS}
\end{equation}
and
\begin{equation}
k_{f}= \left (\frac {3\pi^2}{2}\rho_{nm} \right)^{1/3}.
\end{equation}
In  Eqs. (\ref{t0}) - (\ref{alpha}), the various quantities
characterizing the nuclear matter are the binding energy per nucleon
$B/A$, isoscalar effective mass $m^*/m$, nuclear matter incompressibility
coefficient $K_{nm}$, symmetry energy coefficient $J=S(\rho=\rho_{nm})$,
the coefficient $L=P(\rho=\rho_{nm})$, enhancement factor $\kappa$
and Landau parameter $G_0^\prime$.  All these quantities are taken at
the saturation density $\rho_{nm}$. It must be pointed out  that the
expression for the parameter $G_0^\prime$ used in Eq. (\ref{t2})
includes the contributions from the spin-density term present in the
Skyrme energy density functional \cite{Bender02}.  So, for consistency,
the HF calculations are also performed  by including the contributions
from the spin density.  Once, $T_0$ is known, $T_s$ can be calculated
for a given value of the surface energy $E_s$ as \cite{Margueron02},

\begin{equation}
\label{Es}
E_s=8\pi r_0^2\int_0^{\rho_{nm}} d\rho
\left [\frac{\hbar^2}{36m}-\frac{5}{36}T_0\rho+\frac{1}{8}T_s\rho
-\frac{m^*}{\hbar^2}V_{so}\rho^2\right ]^{1/2}
\left [B(\rho_{nm})/A-B(\rho)/A\right ]^{1/2},
\end{equation}
where, $B(\rho)/A$ is the binding  energy per nucleon given by,
\begin{equation}
\frac{B(\rho)}{A}=-\left [\frac{3\hbar^2}{10m^*}k_f^2+\frac{3}{8}t_0\rho+\frac{1}{16}t_3\rho^{\alpha+1} \right ]
\end{equation}
and,
\begin{eqnarray}
r_0=\left [ \frac{3}{4\pi\rho_{nm}}\right ]^{1/3},\\
V_{\rm so}= \frac{9}{16}{W_0}^2.
\end{eqnarray}
The manner in which Eqs. (\ref{t0}) - (\ref{alpha}) can be used
to evaluate the Skyrme parameters $t_i$, $x_i$  and $\alpha$ is as
follows. At  first, the parameters  $t_0$ and $\alpha$ can be calculated
in terms of $B/A$, $\rho_{nm}$, $K_{nm}$ and $m^*/m$, using Eqs. (\ref{t0})
and (\ref{alpha}). Then, the parameter  $t_3$ can be determined
using Eq. (\ref{t3}).  Next, $T_0$ and $T_s$ can be calculated using
Eqs. (\ref{T0}) and (\ref{Es}), respectively.  Once, the combinations
$T_0$ and $T_s$ of the Skyrme parameters  are known, one can calculate
the remaining parameters in the following sequence, $t_1$, $t_2$, $x_2$,
$x_1$, $x_3$ and $x_0$.

\section{Simulated annealing  based algorithm for the
minimization of $\chi^2$}
\label {sa}

The  simulated annealing method (SAM) is a generalization of a Monte
Carlo technique, based on the  Metropolis algorithm \cite{Metropolis53},
initially developed for examining  the equation of the state of a many body
system. The concept of SAM is based on the manner in which liquids freeze
or metals recrystallize in the process of annealing. In an annealing
process a metal, initially at high temperature and disordered, slowly
cools so that the system at any time is in a  thermodynamic equilibrium. As
cooling proceeds, the system becomes more ordered and approaches a frozen
ground state at zero temperature.

With this brief background, we shall  now implement the SAM to search for 
the global minimum of $\chi^2$ function  as given by Eq. (\ref{chi2}).
One of the crucial key ingredients required to implement  the SAM,  in
the present case, is to specify the lower and the upper limits for each
of the Skyrme parameters. So that the global minimum for the $\chi^2$
is searched  within these limits. However, from the literature (e.g., see
Refs.  \cite{Gomez92,Stone03})  we find that the Skyrme parameters vary
over a wide range. To make the search process  more efficient, we make
use of the fact that most of the Skyrme parameters can be expressed in
terms of the various quantities related to the nuclear matter properties
as described in Sec. \ref{sknm}.  Most of these nuclear matter quantities
are known empirically within $10\% - 20\%$.  For convenience, we define
a vector ${\bf v}$  with 10 components as,
\begin{equation}
{\bf v} \equiv (B/A, K_{nm},  \rho_{nm}, m^*/m, E_s, J, L, \kappa,
G_0^\prime, W_0).
\label{v0}
\end{equation}
Once the vector ${\bf v}$ is known we can calculate the values of all
the Skyrme parameters as discussed in Sec. \ref{sknm}.  We also define
the vectors ${\bf v}_0$, ${\bf v}_1$ and $\bf d$. The vector ${\bf v}_0$
and ${\bf v}_1$ contains the lower and the upper limits of each of the
components of the vector ${\bf v}$. The vector ${\bf d}$ represents the
maximum displacement allowed in a single step for the components of the
vector ${\bf v}$.  We implement the SAM algorithm  using the following
basic  steps,
\begin{itemize}
\item[(i)] We start with a guess value for the vector ${\bf v}$ and
calculate $\chi^2$ (say, $\chi^2_{\rm old}$) using Eq. (\ref{chi2}) for
a given set of the experimental data and the corresponding HF results
together with the theoretical errors.

\item[(ii)] We generate randomly a new set of Skyrme parameters using
the following steps.  First,  we use a uniform random number to  select a
component $v_r$  of the vector ${\bf v}$.  Second,  the randomly selected
component $v_r$ is then assigned a new  value,
\begin{equation}
v_r \rightarrow v_r + \eta d_r,
\end{equation}
where  $\eta$ is a uniform random number which lies within  the range of
$-1$ to $+1$. The second step is repeated until the new value of $v_r$
is found within its allowed limits defined by ${\bf v}_0$ and ${\bf
v}_1$. We use this modified ${\bf v}$ to generate a new set of Skyrme
parameters.  It may be noted that a change in the value of a component
of the vector ${\bf v}$ may lead to changes in the  values of several
Skyrme parameters. For example, a  change in the value of $K_{nm}$
will alter the values of the Skyrme parameters $t_0$, $t_3$ and $\alpha$.

\item[(iii)] The newly generated set of the Skyrme parameters is accepted
by using the Metropolis algorithm as follows.  We calculate  the quantity,
\begin{equation}
{\cal {P}}(\chi^2) = e^{(\chi^2_{\rm old} - \chi^2_{\rm new})/T},
\end{equation} where $\chi^2_{\rm new}$ is obtained by using the newly
generated set of the  Skyrme parameters and $T$ is a control parameter
(an effective temperature). The new set of Skyrme parameters is accepted
only  if,
\begin{equation}
{\cal P}(\chi^2) > \beta,
\label{prob}
\end{equation}
where $\beta$  is  a uniform random number which lies between 0 and 1.  If the
new Skyrme parameters are accepted (i.e. Eq. (\ref{prob}) is satisfied), it is
called a "successful reconfiguration".  \end{itemize}

To search for  the global minimum of $\chi^2$ we begin with some
reasonable value of an   effective temperature  $T = T_i$.  For a given $T_i$,
we repeat steps (ii) and (iii) for, say, $100N_p$ reconfigurations, or
for $10N_p$ successful reconfigurations, whichever comes first.  Then,
we reduce the temperature by following a suitable annealing schedule.
One encounters various annealing schedule available in the literature such
as linear, exponential, Boltzmann and Cauchy annealing schedules
\cite{Cohen94}. Among these, the Boltzmann annealing schedule is the
slowest one and the exponential annealing schedule is the fastest one.
In the present work we have employed the Cauchy annealing schedule given
by, \begin{equation} T(k)=T_i/ck \label{Tk} \end{equation} where, $c$
is a constant, which is taken to be unity in the present work, and $k =
1, 2, 3, .....$ is the time index.  We keep on reducing the value of $T$
using Eq. (\ref{Tk}) in the subsequent steps  until the effort to reduce
the value of $\chi^2$ further becomes sufficiently discouraging.

In Table \ref{in1} we list the values of all the components of the
vectors ${\bf v}$, ${\bf v}_0$, ${\bf v}_1$ and ${\bf d}$ used in the
numerical computation. We have varied the components of the vector ${\bf
v}$  over a wide range. The values of the maximum displacement as defined
by the components of $\bf d$ are so chosen that the corresponding
component of the vector ${\bf v}$ can be varied over the entire range
given by the vectors ${\bf v_0}$ and ${\bf v_1}$, within the adopted
number of reconfigurations. We have carried out
several sample runs and found that $T_i = 1.25$ along with the Cauchy
annealing schedule yields reasonable values of the Skyrme parameters.
We must mention here that the range for the quantities  $L$, $\kappa$
and $G_0^\prime$ as given in Table \ref{in1} are so chosen that they
vary within acceptable limits \cite{Agrawal04}.

\section{Experimental data and some constraints}
\label{bain}
In  this section we discuss our selection  of the experimental data
and the corresponding theoretical errors adopted  in the $\chi^2$ fit,
Eq. (\ref{chi2}), to the HF results.  In Table \ref{in4} we
summarize our choice of the experimental data. It must be noted that in
addition to the typically used data on the binding energy, charge radii
and spin-orbit splitting, we also include in our fit the experimental
data for the radii of valence neutron orbits and the breathing mode
energies of several nuclei. All of these experimental data are
taken from Refs. \cite{Audi03, Otten89,vries87,Kalantar88,Platchkov88,
Trache96,Youngblood99}.  For the binding energy we use in our fit the
error of 1.0 MeV except for the $^{100}$Sn nuclei. The
binding energy for the $^{100}$Sn nucleus is determined
from systematics and are expected to have large errors. Thus, we
assign them a  theoretical  error of 2.0 MeV. For the charge rms radii we use
the theoretical error of 0.02 fm except for the case of $^{56}$Ni nucleus. The 
charge rms radii for the $^{56}$Ni nucleus is obtained from systematics
and we use the theoretical error of 0.04 fm. We consider in our fit
the experimental data for the spin-orbit splittings for  the $2p$ neutrons
and protons in the  $^{56}$Ni nucleus and the rms radii for the $1d_{5/2}$ and
$1f_{7/2}$ neutron  orbits in $^{17}$O and $^{41}$Ca nuclei, respectively.
We use \cite{Trache96},
\begin{eqnarray} \epsilon(2p_{1/2}) - \epsilon(2p_{3/2})
= \left \{ \begin{array}{cc} 1.88 \>\text{MeV}& \qquad\qquad
\text{Neutrons}\\ 1.83 \>\text{MeV}&
\qquad\qquad \text{Protons}\\ \end{array} \right .  
\end{eqnarray}
where, $\epsilon$ is the "bare" single-particle energy obtained by
unfolding the experimental data for the energy levels in $^{57}$Ni
and $^{57}$Cu nuclei by appropriately accounting for the coupling to
excitations of the core. 
Of course, it is more appropriate to use the splitting of high $l$ orbits
in a heavy nucleus (e.g., $^{208}$Pb nucleus) to determine the strength of
the spin-orbit interaction. But, to the best of our knowledge, unlike for
the $^{56}$Ni nucleus  the bare single-particle energies for the heavier
nuclei are not available.  For the rms radii of the valence neutron
orbits in $^{17}$O and $^{41}$Ca  nuclei we use $r_v(\nu 1d_{5/2})=3.36$
fm and $r_v(\nu 1f_{7/2}) = 3.99$ fm, \cite{Kalantar88,Platchkov88}
respectively. The theoretical error taken for the spin-orbit splitting
data is 0.2 MeV and for the rms radii for the valence neutron orbits
we use the experimental error of 0.06 fm. We must point out that the
choice of the theoretical error on the rms radii for the valence neutron
orbits is due to the large uncertainties associated with their extraction
from the experimental measurements. To be consistent with the way these
valence neutron radii are determined, we do not include the center of
mass correction to these data.  The experimental data for the breathing
mode constrained energies $E_0$ included in our fit are 17.81, 15.90,
15.25 and 14.18 MeV  for the  $^{90}$Zr, $^{116}$Sn, $^{144}$Sm and
$^{208}$Pb nuclei \cite{Youngblood99}, respectively, with the theoretical
error taken to be 0.5 MeV for the  $^{90}$Zr nucleus and 0.3 MeV for
the other nuclei. We also include the critical density  $\rho_{cr}$ in
the fit assuming a value of $2.5\rho_0$ with an error of $0.5\rho_0$.
Further the values of the Skyrme parameters are constrained by requiring
that  (i) $P \geqslant 0$ for $\rho \leqslant 3\rho_0$, (ii) $\kappa
=0.1 - 0.5$ and (iii) $G_0^\prime \geqslant 0$ at $\rho = \rho_0$.

\section{Results and discussions}
\label{res}
In the preceding sections we have described in detail the implementation
of the SAM based algorithm to fit the values of the Skyrme parameters to
a given set of the experimental data considered in this work.  We have
carried out two different fits. These fits are carried out by using the
same set of  experimental data along with  some constraints as discussed
in Sec. \ref{bain}. We name these fits as,

\begin{itemize} 
\item[1.] KDE0;  only the direct Coulomb term in the form of Eq. (\ref{Hdir}) is included.
\item[2.] KDE; the direct as well as the exchange Coulomb terms are included
(Eqs. (\ref{Hcoul})-(\ref{Hex})).
\end{itemize}
The CM corrections to the total binding energy, Eqs. (\ref{ecm-osc}) and
(\ref{hw}), and the charge rms radii, Eqs. (\ref{rpav}) and (\ref{rc0}),
are carried out using the schemes described in Sec. \ref{cmc}.

We shall first consider some technical aspects required to implement
the SAM. As it is evident from Sec. \ref{sa}, there are two crucial
ingredients, namely, (i) initial value for the control parameter $T =
T_i$ and (ii) annealing schedule which determines the subsequent value
for $T$.  These ingredients essentially controls the computer time and
the quality of the final fit. If one starts with a smaller value for $T_i$
and/or uses a faster annealing schedule, one may not be able to  hit the
global minimum of the objective function  and rather get trapped in one
of the local minima.  In the present work we have employed the Cauchy
annealing schedule.  We have carried out several trial calculations and
find that $T_i = 1.25$,  along with the Cauchy's annealing schedule
as given by Eq. (\ref{Tk}),  yields reasonable values for the best
fit parameters.  To validate the present approach we carried out
the following checks.  Starting with the final values of the Skyrme
parameters obtained using the SAM, we attempted to minimize further
the value of $\chi^2$ using the Levenberg-Marquardt (LM)  method \cite{NR92}
as conventionally used.  But, we found no further decrease in the value
of the $\chi^2$.  As an illustration, we plot in Fig. \ref{t_chi} the
average value $\langle \chi^2\rangle_{\scriptstyle T}$ as an inverse
function of the control parameter $T$ for the KDE0 case.  The curves
labeled $v$ and $v_1$ represent the results obtained from two different
choices of the starting values for the Skyrme parameters. The initial
value of the Skyrme parameters for the curve labeled $v$ (solid line)
and $v_1$ (dashed line) are  obtained using the set of values given in the
second and fourth  columns of Table \ref{in1}, respectively.  The value of
$\langle \chi^2\rangle_T$ is obtained by averaging over all the successful
reconfigurations for a given $T$.  We see from Fig. \ref{t_chi} that
the value of $\langle \chi^2\rangle_T$ show a remarkable decrease at
initial stages and then oscillates before saturating to a minimum value
for $T\leqslant 0.05$.  The value  of $\chi^2$  at lower $T$  is more
or less independent of the starting values for the Skyrme parameters.
In Fig \ref{t_dchi} we have displayed the variation of $\Delta\chi^2_T
= \langle (\chi^2 - \langle\chi^2\rangle)^2\rangle_T$ as an inverse
function of $T$.  We see that the fluctuations in the value of $\chi^2$
is large for larger values of $T$. As $T$  decreases, fluctuations in
the value of $\chi^2$ also decrease rapidly. This means that  initial
value for $T$ should not be too small, because, at smaller $T$ it is
less likely  to jump from a configuration with lower value of $\chi^2$
to one having higher value. As a result, one may get trapped in a local
minima.  In Table \ref{skm_v01} we give the values of the parameters for
the KDE0 interaction at the minimum value of the $\chi^2$ obtained from
different choices for the starting  values for the Skyrme parameters.
It is interesting to note that not  only the final value of the $\chi^2$
is less sensitive to the  choice of the initial parameters, but,
also the resulting Skyrme parameters are also quite close.  In what
follows, we shall present the results for the KDE0 and KDE interactions.
The starting (or guess) values for he Skyrme parameters used to generate
these interactions are obtained from the nuclear  matter quantities
given in the second column of the Table \ref{in1}.

In Table \ref{nm_par} we give the values for the  various quantities
characterizing the nuclear matter obtained at the minimum value of
the $\chi^2$.  We also note that the values of  all the nuclear matter
properties for the KDE0 and KDE Skyrme interaction are closer to those
obtained for the SLy7 interaction.  However, it is worth mentioning
that the values of the $K_{nm}$ and $m^*/m$ for both the interactions
generated here emerge from the fit, unlike the SLy type interactions
where the values for these quantities were kept fixed.  In our fits,
the values of the $K_{nm}$ and $m^*/m$ are mainly constrained by
the inclusion of the experimental data on breathing mode energy
and the value of critical density $\rho_{cr}=2.5\rho_0\pm 0.5\rho_0$
\cite{Margueron02,Agrawal04}.  In the last row of this table we give the
values of $\chi^2$ at the minimum.  For the sake of completeness, we list
in Table \ref{skm_par} the values of the Skyrme parameters obtained in
the fits.  One can easily calculate the values of these Skyrme parameters
using the various nuclear matter quantities given in Table \ref{nm_par}
as described in Sec. \ref{sknm}.  In Table \ref{skm_par} we also give in
parenthesis the values of the  standard deviations for the Skyrme
parameters.    Since, within the SAM algorithm one can not calculate
these standard deviations in a straight forward manner, we resort to some
alternative  approach.  We have determined the values of the standard
deviations on the parameters for the KDE0 and KDE interactions using
the LM method. The LM method requires two inputs, namely, set of the
experimental data and the starting values of the interaction parameters.
The set of experimental data is taken to be exactly the same as the one
used to generate the KDE0 and KDE interactions. The starting values of
the interactions parameters used are the ones obtained using SAM for
the KDE0 and KDE interactions.

In Table \ref{in2} we present our results for the deviation $\Delta B =
B^{exp} - B^{th}$ for the values of the binding energy obtained from
the newly generated KDE0 and KDE interactions. Similar
deviations $\Delta r_{ch}$  for the charge rms radii are presented
in Table \ref{in3}.  For comparison, in the last columns of these
Tables we give the values of $\Delta B$ and $\Delta r_{ch}$ for the
SLy7 interaction,  taken from Ref. \cite{Chabanat98}.  One can easily
verify from Table \ref{in2} that the magnitude of the deviations for
the binding energy for most of the cases is much less than $0.5 \%$
in case of KDE0 interaction. The KDE interaction yields larger error in
the values of the binding energy ($\sim 0.6 - 1.0\%$)  for the $^{16}$O,
$^{48}$Ni and $^{100}$Sn nuclei.  We would also like to remark here that
in determining the SKX interaction, the binding energy for the   $^{56}$Ni
nucleus was not considered in the fit and that for  the $^{100}$Sn nucleus
was included in the fit with the theoretical error of 1.0 MeV. We find
that if one attempt to do so, the binding energy for the $^{56}$Ni becomes
off by more than 3 MeV.  We see from Table \ref{in3} that, except for the
$^{16}$O and $^{48}$ca nuclei, the deviations in the values of the charge
rms radii for the KDE0 interaction is less than 0.5\%.  In addition to the
binding energy and the charge rms radii of the nuclei used in our fits,
we have also considered a few more experimental data as discussed in
Sec. \ref{bain}. In Tables \ref{extra_data} and \ref{e0_tab} we present
our results for these additional quantities.  The values of $\rho_{cr}$
is greater than $2\rho_0$. The values for the radii of valence neutron
orbits  and the spin-orbit splittings considered in our fits are quite
reasonable for all the interactions considered here.  It can be seen
from Table \ref{e0_tab} that our fit to the breathing mode constrained
energies are overall in reasonable agreement with the corresponding
experimental data.

We now consider our results for the binding energy difference
between the $^{48}$Ca and $^{48}$Ni mirror nuclei.  One may verify
from Table \ref{in2} that the binding energy difference $B(^{48}{\rm
Ca}) - B(^{48}{\rm Ni})$ = 67.23 and  64.02 MeV for the KDE0 and KDE
interactions, respectively, compared to the experimental value of
68.85 MeV.  We would also like to add that the said difference for the
SKX interaction is 66.3 MeV which is about 1.0 MeV lower than our most
realistic KDE0 interaction.  On the other hand,  most of the Skyrme
interactions which include the contribution from the exchange Coulomb
term yield $B(^{48}{\rm Ca})-B(^{48}{\rm Ni})\approx 63$ MeV,  which is
about 6 MeV lower than the corresponding experimental value.

We present in Table \ref{rn-rp} our results  for the neutron skin, $r_n
- r_p$, the difference between the rms radii for the point neutrons
and protons density distributions, for the KDE0 and KDE interactions.
We compare in Tables \ref{Ca40} and \ref{Pb208} the values of the
single-particle energies with the available experimental data for the
$^{40}$Ca and $^{208}$Pb nuclei \cite{Bohr69,James69}, respectively.
We find that the single-particle energies, for the occupied states near
the fermi-energy compare reasonably well with the experimental ones.
We would like to remark here  that the HF approach alone is not expected
to reproduce the experimental single-particle energies and there fore
we have not included them in our fit.

Finally, we consider  the behavior of  the symmetry energy coefficient
$S(\rho)$ for densities relevant to the study of neutron stars. It is
well known \cite{Kutschera94,Kutschera00} that the values of $S(\rho)$
and the resulting EOS for pure neutron matter at higher densities ($\rho >
2\rho_0$)  are crucial in understanding the various properties of
neutron star. For example, the proton fraction at any density depends
strongly on the value of $S(\rho)$ at that density,  which in turn
affects the chemical compositions as well as the cooling mechanism of
the neutron star \cite{Lattimer91}.  Yet, no  consensus is reached for
the density dependence of $S(\rho)$.  We display in Fig. \ref{J_rho},
our results for the variation of the symmetry energy $S$ as a function
of the nuclear matter density $\rho$. We see for the  KDE0 and KDE
interactions that the  value of $S$ increases with density for $\rho <
3\rho_0$.  All of these interactions are quite suitable for  modeling
the neutron star with masses close to the canonical one \cite{Stone03},
because, they yield $S > 0$ for $\rho < 4\rho_0$.  In Fig. \ref{eos_neu}
we plot the EOS for the pure neutron matter resulting from the KDE0
and KDE interactions and compare them with the ones obtained for SLy7
interaction and the realistic UV14+UVII model \cite{Wiringa88}. It is
striking to note that our results for the  KDE0 and KDE interactions
are in harmony with the EOS for the UV14+UVII model, though, unlike the
SLy7 interaction we did not include in our fit the neutron matter EOS
of the realistic UV14+UVII interaction. This seems to be due to the
constrain imposed on the quantity $P$, which is  related to the slope
of the symmetry energy coefficient (see Eq. (\ref{P})).

\section{Conclusions}
\label{conc}
We have implemented the simulated annealing method to fit the values
of the parameters of the Skyrme interaction   of Eq. (\ref{v12}) by
searching for the global minimum  in the hyper surface of the $\chi^2$
function, Eq. (\ref{chi2}). To demonstrate the applicability of this
method we have fitted the values of the Skyrme parameters to an extensive
set of experimental data together with a few additional constraints.
Our experimental data set consists of the binding energies for 14 nuclei
ranging from the normal to exotic (proton or neutron rich) ones,
charge rms radii for 7 nuclei, spin-orbit splittings for the $2p$ proton and
neutron orbits of the  $^{56}$Ni nucleus and  rms radii for $1d_{5/2}$
and $1f_{7/2}$ valence neutron orbits in the $^{17}$O and $^{41}$Ca nuclei,
respectively. We also include in the fit the critical density $\rho_{cr}$
determined from   the stability conditions for the  Landau parameters.
The additional constraints imposed on the Skyrme parameters are (i)
the quantity $P = 3\rho \frac{dS}{d\rho}$, directly related to the slope
of the symmetry energy $S$, must be positive for densities up to $
3\rho_0$; a condition imposed by the neutron star models \cite{Stone03},
(ii) the enhancement factor $\kappa$, associated with the TRK sum rule
for the isovector giant dipole resonance, should lie in the range of
$0.1 - 0.5$ and (iii) the Landau parameter $G_0^\prime$, crucial for the
spin properties of finite nuclei and nuclear matter, should be
positive at $\rho = \rho_0$.

Using these experimental data along with the additional constrains, we
have carried out two different fits named as KDE0 and KDE, as described
in Sec. \ref{res}.  The corrections to the binding energy and charge
rms radii due to the center of mass motion were performed using  simple
but consistent schemes.  The nuclear matter properties for both 
interactions proposed in the present work  are obtained directly from
the fit.  The selection of the experimental data in conjugation with some
constraints ensures that these interactions  can be used to study the
bulk ground state  properties  of nuclei ranging from the stable to the
ones  near the proton and neutron drip lines, as well as the properties
of neutron stars.  The interactions obtained in the present work
encompasses the merits of the SKX and SLy type of Skyrme interactions.

Before closing, we would like to mention that the method as well as the
fitting strategy presented in this work can be improved in several ways.
The SAM is a very adaptive approach and therefore it offers a significant
scope for further improvement.  For example, in the present work
we jump from one configuration to another by randomly selecting a
component of the vector ${\bf v}$ as defined by Eq. (\ref{v0}). This
selection was done using a uniform random number. However, one can think
of performing random selection of a component of ${\bf v}$ by assigning
a more plausible weight factors to these components. One can also try
out different annealing schedules to determine the rate of cooling. In
the present work we employed the  Cauchy annealing schedule  which yields a
faster cooling rate than that of the Boltzmann schedule, but, a slower
rate than  the exponential annealing schedule.  The effects on the
binding energy and radii due to the correlations beyond mean-field
\cite{Bertsch83,Shlomo88,Bender04} can be included in the fit. These effects are
in particular important for the light nuclei.  One may also  include in the
spin-orbit splitting the contributions due to the electromagnetic
spin-orbit interaction \cite{Trache96} and modify the spin-orbit
interaction by using the form proposed by Sagawa in Ref.  \cite{Sagawa01}.
Last but not least, one may also include the experimental data on the
giant dipole and quadrupole resonances while fitting the Skyrme parameters
in addition to the breathing mode energy, as was done in the present work.

\begin{acknowledgments}
This work was supported in part by the US Department of Energy under grant
\# DOE-FG03-93ER40773 and the National Science Foundation under grant
$\#$PHY-0355200. 
\end{acknowledgments}
\newpage

\newpage
\begin{figure} 
\noindent{\bf Figure captions}
\caption{\label{t_chi} Variation of the average value of chi-square,
$\langle\chi^2\rangle_T$,  as  a function of the inverse of the control
parameter $T$ for the KDE0 interaction for the two different choices of the
starting parameters (see text for detail).}

\caption{\label{t_dchi} Variation of the fluctuations $\Delta
\chi^2_T$ in the value of $\chi^2$  as a function of $1/T$
for the KDE0 interaction for the two different choices of the
starting parameters (see text for detail).}

\caption{\label{J_rho} Variation of the symmetry energy coefficient
$S(\rho)$
as a function of the nuclear matter density $\rho$.}

\caption{\label{eos_neu} Energy per particle for pure neutron matter
$E^{(n)}/A$ as a function of density. Results for  the two newly generated
Skyrme interactions KDE0 and KDE are compared with those obtained
for the SLy7 Skyrme force and the realistic UV14+UVII model  of Wiringa
et al.  \cite{Wiringa88}. }

\end{figure}

\newpage
\vspace*{4.0 true in}
\begin{table}[p]
\caption{\label{in1} Values of the components of the vectors ${\bf v}$,
${\bf v}_0$, ${\bf v}_1$ and ${\bf d}$ used for implementing  the
SAM based algorithm for searching the global minimum of $\chi^2$ . The
vector ${\bf v}$ initializes the value of $\chi^2$, whereas, ${\bf v}_0$
and ${\bf v}_1$ limits the search space for the Skyrme parameters. The
components of the  vector ${\bf d}$ correspond to  the maximum
displacements allowed for the reconfiguration.}

\begin{ruledtabular}
\begin{tabular}{|cdddd|}
\multicolumn{1}{|c}{}&
\multicolumn{1}{c}{${\bf v}$}&
\multicolumn{1}{c}{${\bf v}_0$}&
\multicolumn{1}{c}{${\bf v}_1$}&
\multicolumn{1}{c|}{${\bf d}$}\\
\hline
$B/A $(MeV)  &16.0 &17.0&  15.0&  0.40\\
$K_{nm}$(MeV) &230.0  &200.0&  300.0&  20.0\\
$\rho_{nm}$(fm$^{-3}$) &0.160  &0.150& 0.170& 0.005\\
$m^*/m$ &0.70 &0.60&  0.90& 0.04\\
$E_s$(MeV) & 18.0    &17.0&  19.0&  0.3\\
$J$(MeV) & 32.0    &25.0&  40.0&  4.0\\
$L$(MeV) & 47.0    &20.0&  80.0& 10.0\\
$\kappa$ & 0.25    &0.1&  0.5&  0.1\\
$G_0^\prime$     & 0.08    &0.00 & 0.40&  0.10\\
$W_0$ (MeV.fm$^5$)    &  120.0   &100.0&  150.0 & 5.0\\
\end{tabular}
\end{ruledtabular}
\end{table}
\begin{table}
\caption{ \label{in4} 
Selected  experimental  data  for the binding energy $B$, charge rms 
radius $r_{ch}$,  rms radii of valence neutron orbits $r_v$,  spin-orbit
splitting S-O,  breathing mode constrained energy $E_0$ and critical
density $\rho_{cr}$   used in the  fit to determine the parameters of
the Skyrme interaction.}
\begin{ruledtabular}
\begin{tabular}{|ccc|}
Properties& Nuclei& Ref.\\
\hline
$B$ & $^{16, 24}$O,  $^{34}$Si, $^{40,48}$Ca, $^{48,56,68,78}$Ni,
 $^{88}$Sr, $^{90}$Zr, $^{100,132}$Sn, $^{208}$Pb& \cite{Audi03}\\
$r_{ch}$& $^{16}$O, $^{40,48}$Ca, $^{56}$Ni, $^{88}$Sr, $^{90}$Zr,
$^{208}$Pb& \cite{Otten89,vries87}\\
$r_v(\nu 1d_{5/2})$&$^{17}$O  & \cite{Kalantar88}\\
$r_v(\nu 1f_{7/2})$&$^{41}$Ca & \cite{Platchkov88}\\
S-O& $2p$ orbits in $^{56}$Ni  & \cite{Trache96}\\
$E_{o}$& $^{90}$Zr, $^{116}$Sn, $^{144}$Sm, $^{208}$Pb & \cite{Youngblood99} \\
$\rho_{cr}$ &   nuclear matter & see text\\
\end{tabular}
\end{ruledtabular}
\end{table}
\newpage
\begin{table}[p]
\caption{\label{skm_v01} 
Comparison of the parameters for the KDE0 interaction  at
the minimum value of $\chi^2$ obtained from different choices  for the
starting values of the Skyrme parameters.}
\begin{ruledtabular}
\begin{tabular}{|cdd|}
\multicolumn{1}{|c}{Parameter}&
\multicolumn{1}{c}{KDE0(${\bf v}$)}&
\multicolumn{1}{c|}{KDE0(${\bf v}_1$)}\\
\hline
  $t_0$(MeV$\cdot$fm$^3$)  & -2526.51&-2553.08  \\
  $t_1$(MeV$\cdot$fm$^5$)  & 430.94& 411.70 \\
  $t_2$(MeV$\cdot$fm$^5$)  & -398.38&-419.87  \\
  $t_3$ (MeV$\cdot$fm$^{3(1+\alpha)}$) & 14235.5&  14603.6\\
  $x_0$  &  0.7583& 0.6483 \\
  $x_1$  & -0.3087& -0.3472 \\
  $x_2$  &  -0.9495&-0.9268 \\
  $x_3$  &  1.1445& 0.9475 \\
  $W_0$(MeV$\cdot$fm$^5$)  & 128.96& 124.41 \\
  $\alpha$  &  0.1676& 0.1673 \\
\hline
\end{tabular}
\end{ruledtabular}
\end{table}
\newpage
\begin{table}[p]
\caption{\label{nm_par} Nuclear matter properties for the 
KDE0 and KDE  interactions at the $\chi^2 = \chi^2_{min} $.}
\begin{ruledtabular}
\begin{tabular}{|cddd|}
\multicolumn{1}{|c}{Parameter}&
\multicolumn{1}{c}{KDE0}&
\multicolumn{1}{c}{KDE}&
\multicolumn{1}{c|}{SLy7}\\
\hline
  $B/A$ (MeV) & 16.11  & 15.99  & 15.92\\
  $K_{nm}$(MeV)  & 228.82  & 223.89 &229.7\\
  $\rho_{nm}$  &   0.161  &   0.164  &0.158\\
  $m^*/m $  &   0.72  &   0.76 &0.69\\
  $ E_s$  (MeV)&  17.91  &   17.98  &17.89\\
  $J $ (MeV) &  33.00  &  31.97  &31.99\\
  $L$ (MeV) &  45.22  & 41.43  &47.21\\
  $\kappa$  &   0.30  &   0.16 &0.25\\
  $G^\prime_0$  &   0.05  &   0.03   &0.04\\
$ \chi^2_{min}$ & 1.3&2.2 & \\
\end{tabular}
\end{ruledtabular}
\end{table}

\newpage
\begin{table}[p]
\caption{\label{skm_par} The values of the Skyrme parameters for KDE0
and KDE interactions obtained by minimizing the $\chi^2$.  For the sake
of comparison we have also listed the values of the parameters for the
SLy7 interaction.  The values in Parenthesis are the standard deviation for
the corresponding Skyrme parameters. }
\begin{ruledtabular}
\begin{tabular}{|cccc|}
\multicolumn{1}{|c}{Parameter}&
\multicolumn{1}{c}{KDE0}&
\multicolumn{1}{c}{KDE}&
\multicolumn{1}{c|}{SLy7}\\
\hline
  $t_0$(MeV$\cdot$fm$^3$)  & -2526.51\ (140.63) & -2532.88\ (115.32) & -2482.41\\
  $t_1$ (MeV$\cdot$fm$^5$) &  430.94\ (16.67)  &403.73\ (27.63)  &457.97\\
  $t_2$ (MeV$\cdot$fm$^5$) & -398.38\ (27.31)  &-394.56\ (14.26)  &-419.85\\
$t_3$(MeV$\cdot$fm$^{3(1+\alpha)}$)  & 14235.5\ (680.73)  & 14575.0\ (641.99)  &13677.0\\
  $x_0$  &    0.7583\ (0.0655)  &  0.7707\ (0.0579)  &0.8460\\
  $x_1$  &  -0.3087\  (0.0165)  & -0.5229\ (0.0298)  &-0.5110\\
  $x_2$  &   -0.9495\  (0.0179)  & -0.8956\ (0.0270)  &-1.0000\\
  $x_3$  &    1.1445\  (0.0862)  &  1.1716\ (0.0767)  &1.3910\\
  $W_0$ (MeV$\cdot$fm$^5$) &  128.96\  (3.33)  &128.06\  (4.39)  &126.00\\
  $\alpha$  &    0.1676\  (0.0163)  & 0.1690\  (0.0144) &0.1667\\
\hline
\end{tabular}
\end{ruledtabular}
\end{table}

\newpage
\begin{table}[p]
\caption{\label{in2} Results for the total binding energy $B$ (in MeV)
for several nuclei. The experimental data $B^{exp}$  used to fit the
Skyrme parameters were taken from \cite{Audi03}. The theoretical error
$\sigma$ was taken to be 2.0 MeV for the $^{100}$Sn nucleus and 1.0
MeV for the other nuclei.  In 3rd and 4th columns we give the values
for $\Delta B = B^{exp} - B^{th}$ obtained from our new fits.
The last column contains the values for $\Delta B$ for the SLy7 Skyrme
interaction taken from Ref. \cite{Chabanat98}.}
\begin{ruledtabular}
\begin{tabular}{|cdddd|}
\multicolumn{1}{|c}{}&
\multicolumn{1}{c}{}&
\multicolumn{3}{c|}{$\Delta B=B^{exp} - B^{th}$}\\
\cline{3-5}
\multicolumn{1}{|c}{Nuclei}&
\multicolumn{1}{c}{$B^{exp}$}&
\multicolumn{1}{c}{KDE0}&
\multicolumn{1}{c}{KDE}&
\multicolumn{1}{c|}{SLy7}\\
\hline
  $^{16}$O  & 127.620  &  0.394  &   1.011  & -0.93\\
  $^{24}$O  & 168.384  &   -0.581  &   0.370 &  \\
  $^{34}$Si  & 283.427  &  -0.656  &   0.060 &  \\
  $^{40}$Ca  & 342.050  & 0.005  &   0.252 & -2.85 \\
  $^{48}$Ca  & 415.990  & 0.188  &   1.165 & 0.11\\
  $^{48}$Ni  & 347.136  &-1.437  &  -3.670  & \\
  $^{56}$Ni  & 483.991  &  1.091  &   1.016 &1.71  \\
  $^{68}$Ni  & 590.408  & 0.169  &   0.539 & 1.06 \\
  $^{78}$Ni  & 641.940  & -0.252  &   0.763   & \\
  $^{88}$Sr  & 768.468  &  0.826  &   1.132 & \\
  $^{90}$Zr  & 783.892  &   -0.127  &  -0.200 & \\
  $^{100}$Sn  & 824.800  & -3.664  &  -4.928 &-4.83 \\
  $^{132}$Sn  &1102.850  &  -0.422  &  -0.314  & 0.08 \\
  $^{208}$Pb  &1636.430  & 0.945  &  -0.338  &-0.33 \\
\hline
\end{tabular}
\end{ruledtabular}
\end{table}

\newpage
\begin{table}[p]
\caption{\label{in3} 
Results for the  charge root  mean square (rms)  radii $r_{ch}$
(in fm).  The experimental data used in the fit to determine the values
of the Skyrme parameters are taken from Refs. \cite{Otten89,vries87}.
The theoretical error $\sigma$ taken to be 0.04 fm for the  $^{56}$Ni
nucleus and 0.02 fm for the other nuclei.  In 3rd and 4th columns we give
the values for $\Delta r_{ch} = r_{ch}^{exp} - r_{ch}^{th}$ obtained from
our new fits. The last column contains the values for $\Delta r_{ch} $
for the SLy7 Skyrme interaction taken from Ref. \cite{Chabanat98}.}
\begin{ruledtabular}
\begin{tabular}{|cdddd|}
\multicolumn{1}{|c}{}&
\multicolumn{1}{c}{}&
\multicolumn{3}{c|}{$\Delta r_{ch} = r_{ch}^{exp} - r_{ch}^{th}$}\\
\cline{3-5}
\multicolumn{1}{|c}{Nuclei}&
\multicolumn{1}{c}{$r_{ch}^{exp}$}&
\multicolumn{1}{c}{KDE0}&
\multicolumn{1}{c}{KDE}&
\multicolumn{1}{c|}{SLy7}\\
\hline
  $^{16}$O  &   2.730  & -0.041  &  -0.039 &-0.017\\
  $^{40}$Ca  &   3.490  &0.000  &   0.011 &0.020\\
  $^{48}$Ca  &   3.480  &-0.021  &  -0.008 &-0.015\\
  $^{56}$Ni  &   3.750  &-0.018  &   0.000 &-0.008\\
  $^{88}$Sr  &   4.219  &  -0.002  &   0.019   & \\
  $^{90}$Zr  &   4.258  & -0.008  &   0.013  & \\
  $^{208}$Pb  &   5.500  & 0.011  &   0.041 &0.002\\
\hline
\end{tabular}
\end{ruledtabular}
\end{table}

\newpage
\begin{table}[p]
\caption{\label{extra_data}
Critical density $\rho_{cr}$, rms radii of the valence neutron orbits $r_v$
and spin-orbit splitting (S-O).  The experimental values ( and the
theoretical error $\sigma$) used in the fit to determine the Skyrme
parameters are taken as follows: for the $\rho_{cr}$ we assume a value
of $2.5\rho_0$ ($\sigma=0.5\rho_0$), the values of $r_v$ were  taken from
Ref. \cite{Kalantar88,Platchkov88} ($\sigma=0.06$ fm) and the spin-orbit in
$^{56}$Ni were taken from Ref. \cite{Trache96} ($\sigma = 0.2$ MeV). In
columns $3 - 6$ we give the results obtained from our new fits.}
\begin{ruledtabular}
\begin{tabular}{|cddd|}
\multicolumn{1}{|c}{}&
\multicolumn{1}{c}{Expt.}&
\multicolumn{1}{c}{KDE0}&
\multicolumn{1}{c|}{KDE}\\
\hline
$\rho_{cr}/ \rho_0$ &2.5 &2.5&2.1\\
$r_v (\nu 1d_{5/2}) $(fm) &3.36 & 3.42& 3.41\\
$r_v (\nu 1f_{7/2}) $(fm) & 3.99 &4.05& 4.03\\
$\epsilon _n(2p_{1/2})- \epsilon _n(2p_{3/2})$ (MeV)&1.88& 1.84& 1.81\\
$\epsilon _p(2p_{1/2})- \epsilon _p(2p_{3/2})$ (MeV)&1.83& 1.64& 1.63\\
\hline
\end{tabular}
\end{ruledtabular}
\end{table}

\begin{table}[p]
\caption{\label{e0_tab} Comparison of the breathing mode constrained
energies (in MeV)  obtained for the KDE0 and KDE interactions with
the experimental data.}
\begin{ruledtabular}
\begin{tabular}{|cddd|}
\multicolumn{1}{|c}{Nucleus}&
\multicolumn{1}{c}{Expt.}&
\multicolumn{1}{c}{KDE0}&
\multicolumn{1}{c|}{KDE}\\
\hline
$^{90}$Zr& 17.81& 17.98& 17.91\\
$^{116}$Sn& 15.90& 16.42& 16.36\\
$^{144}$Sm& 15.25& 15.53& 15.47\\
$^{208}$Pb& 14.18& 13.64& 13.60\\
\end{tabular}
\end{ruledtabular}
\end{table}

\newpage

\begin{table}[p]
\caption{\label{rn-rp} 
Results for the  neutron skin, $r_n - r_p$ (in fm), for all the nuclei
considered to obtain the KDE0 and KDE interactions.}
\begin{ruledtabular}
\begin{tabular}{|cdd|}
\multicolumn{1}{|c}{}&
\multicolumn{2}{c|}{$r_n - r_p $ }\\
\cline{2-3}
\multicolumn{1}{|c}{Nuclei}&
\multicolumn{1}{c}{KDE0}&
\multicolumn{1}{c|}{KDE}\\
\hline
$^{16}$O&   -0.031&   -0.025\\
$^{24}$O&    0.510&    0.510\\
$^{34}$SI&    0.189&    0.192\\
$^{40}$CA&   -0.051&   -0.046\\
$^{48}$CA&    0.158&    0.159\\
$^{48}$NI&   -0.282&   -0.274\\
$^{56}$NI&   -0.056&   -0.052\\
$^{68}$NI&    0.175&    0.174\\
$^{78}$NI&    0.287&    0.285\\
$^{88}$SR&    0.095&    0.096\\
$^{90}$ZR&    0.064&    0.065\\
$^{100}$SN&   -0.081&   -0.078\\
$^{132}$SN&    0.220&    0.217\\
$^{208}$PB&    0.160&    0.155\\
\hline
\end{tabular}
\end{ruledtabular}
\end{table}
\begin{table}[p]
\caption{\label{Ca40}Single-particle energies (in MeV)  for $^{40}$Ca
nucleus.}
\begin{ruledtabular}
\begin{tabular}{|ccdd|}
\multicolumn{1}{|c}{Orbits}&
\multicolumn{1}{c}{Expt.}&
\multicolumn{1}{c}{KDE0}&
\multicolumn{1}{c|}{KDE}\\
\hline
\multicolumn{4}{|c|}{Protons}\\
\hline
$1s_{1/2}$&$-50\pm$11&-39.40&-38.21\\
$1p_{3/2}$& --&-26.95& -26.42\\
$1p_{1/2}$&$- 34\pm$6&-22.93& -22.34\\
$1d_{5/2}$& ---& -14.49& -14.51\\
$2s_{1/2}$& $-10.9$& -9.48& -9.66\\
$1d_{3/2}$& $-8.3$&-7.59& -7.53\\
 & && \\
$1f_{7/2}$&$-1.4$&-2.38  &-2.76  \\
\hline
\multicolumn{4}{|c|}{Neutrons}\\
\hline
$1s_{1/2}$&--&-47.77& -46.13\\
$1p_{3/2}$& --& -34.90& -33.92\\
$1p_{1/2}$& --& -30.78& -29.73\\
$1d_{5/2}$& --& -22.08& -21.66\\
$2s_{1/2}$& $-18.1$& -17.00& -16.78\\
$1d_{3/2}$&$-15.6$& -14.97& -14.48\\
& & &\\
$1f_{7/2}$& $-8.32$&-9.60& -9.58\\
$2p_{3/2}$& $-6.2$& -4.98& -5.15\\
\end{tabular}
\end{ruledtabular}
\end{table}
\newpage
\begin{table}[p]
\caption{\label{Pb208}Single-particle energies (in MeV)  for $^{208}$Pb.}
\begin{ruledtabular}
\begin{tabular}{|ccdd|}
\multicolumn{1}{|c}{Orbits}&
\multicolumn{1}{c}{Expt.}&
\multicolumn{1}{c}{KDE0}&
\multicolumn{1}{c|}{KDE}\\
\hline
\multicolumn{4}{|c|}{Protons}\\
\hline
$1g_{9/2}$&  $-15.43$&-17.85& -17.34\\ 
$1g_{7/2}$& $-11.43$& -13.77 &-13.39\\
$2d_{5/2}$& $-9.70$& -11.37& -11.23\\
$1h_{11/2}$& $-9.37$& -9.87& -9.68\\
$2d_{3/2}$& $-8.38$& -9.43& -9.30\\
$3s_{1/2}$& $-8.03$& -8.67& -8.62\\
 & && \\
$1h_{9/2}$& $-3.77$& -4.00& -3.99\\
$2f_{7/2}$& $-2.87$& -2.78& -3.00\\
$1i_{13/2}$& $-2.16$&-1.62& -1.72\\
$3p_{3/2}$& $-0.95$&0.60& 0.26\\
$2f_{5/2}$& $-0.47$&-0.19& -0.42\\
\hline
\multicolumn{4}{|c|}{Neutrons}\\
\hline
$1h_{9/2}$&$-10.85$&-12.39& -12.24\\
$2f_{7/2}$& $-9.72$&-11.60& -11.64\\
$1i_{13/2}$& $-9.01$&-9.33& -9.20\\
$3p_{3/2}$& $-8.27$&-8.67& -8.77\\
$2f_{5/2}$& $-7.95$&-8.59& -8.64\\
$3p_{1/2}$& $-7.38$&-7.54& -7.65\\
 & && \\
$2g_{9/2}$& $-3.94$& -2.86& -3.06\\
$1i_{11/2}$& $-3.15$&-1.65&-1.69\\
$1j_{15/2}$& $-2.53$& -0.41& -0.43\\
$3d_{5/2}$& $-2.36$& -0.43& -0.64\\
$4s_{1/2}$& $-1.91$&0.08& -0.08\\
$2g_{7/2}$& $-1.45$&0.38& 0.20\\
$3d_{3/2}$& $-1.42$&0.56& 0.40\\
\end{tabular}
\end{ruledtabular}
\end{table}
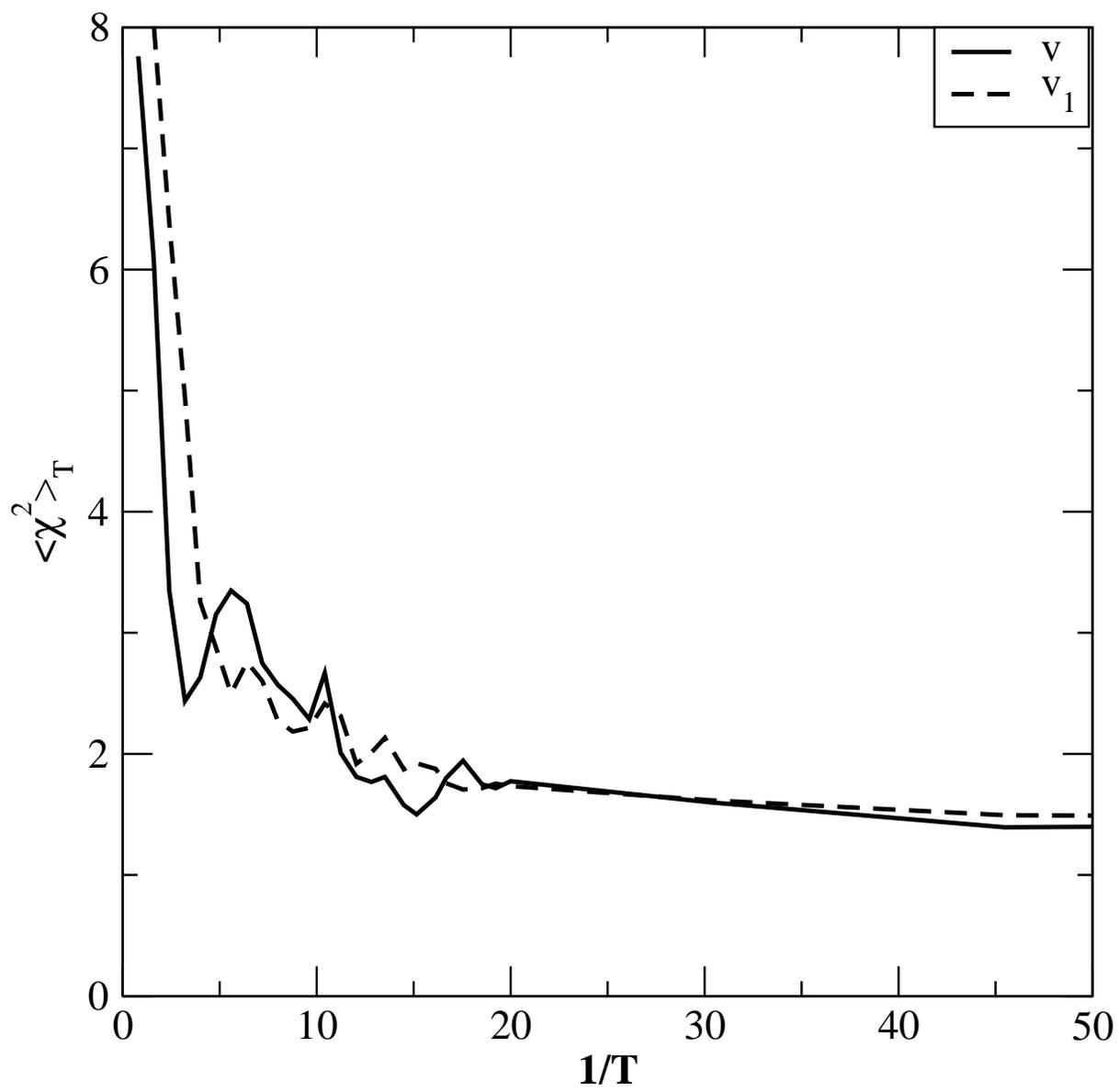
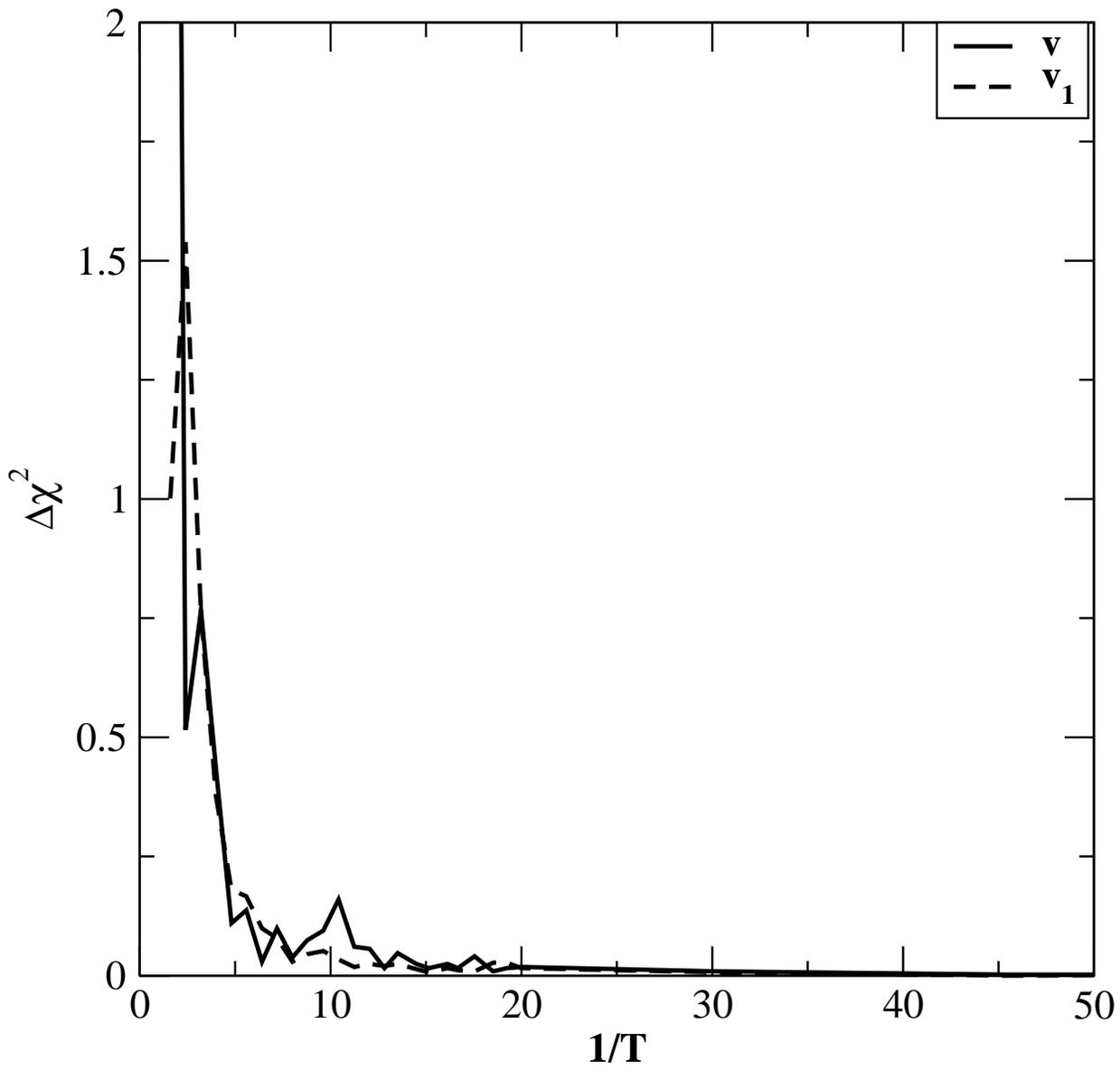
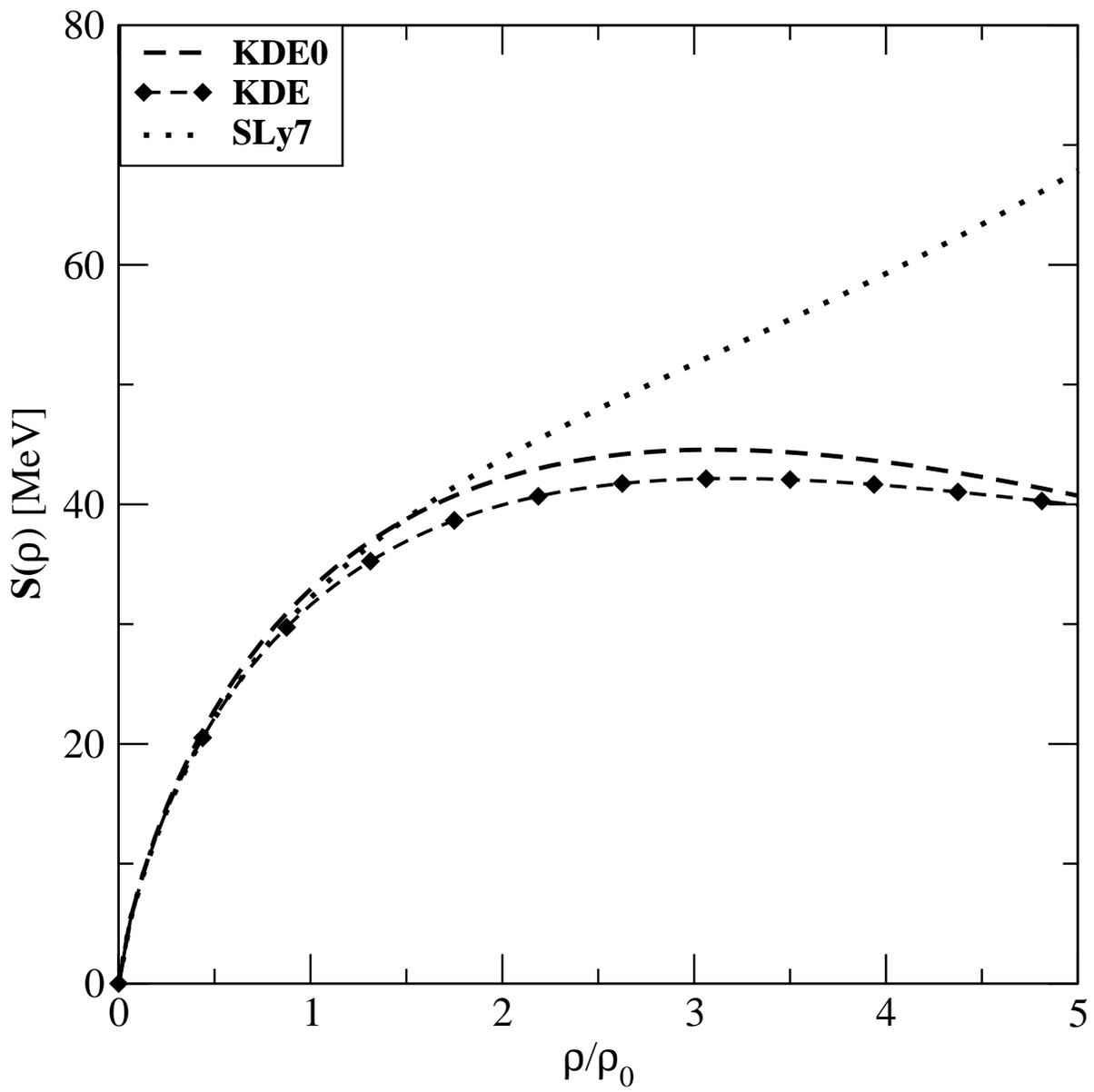
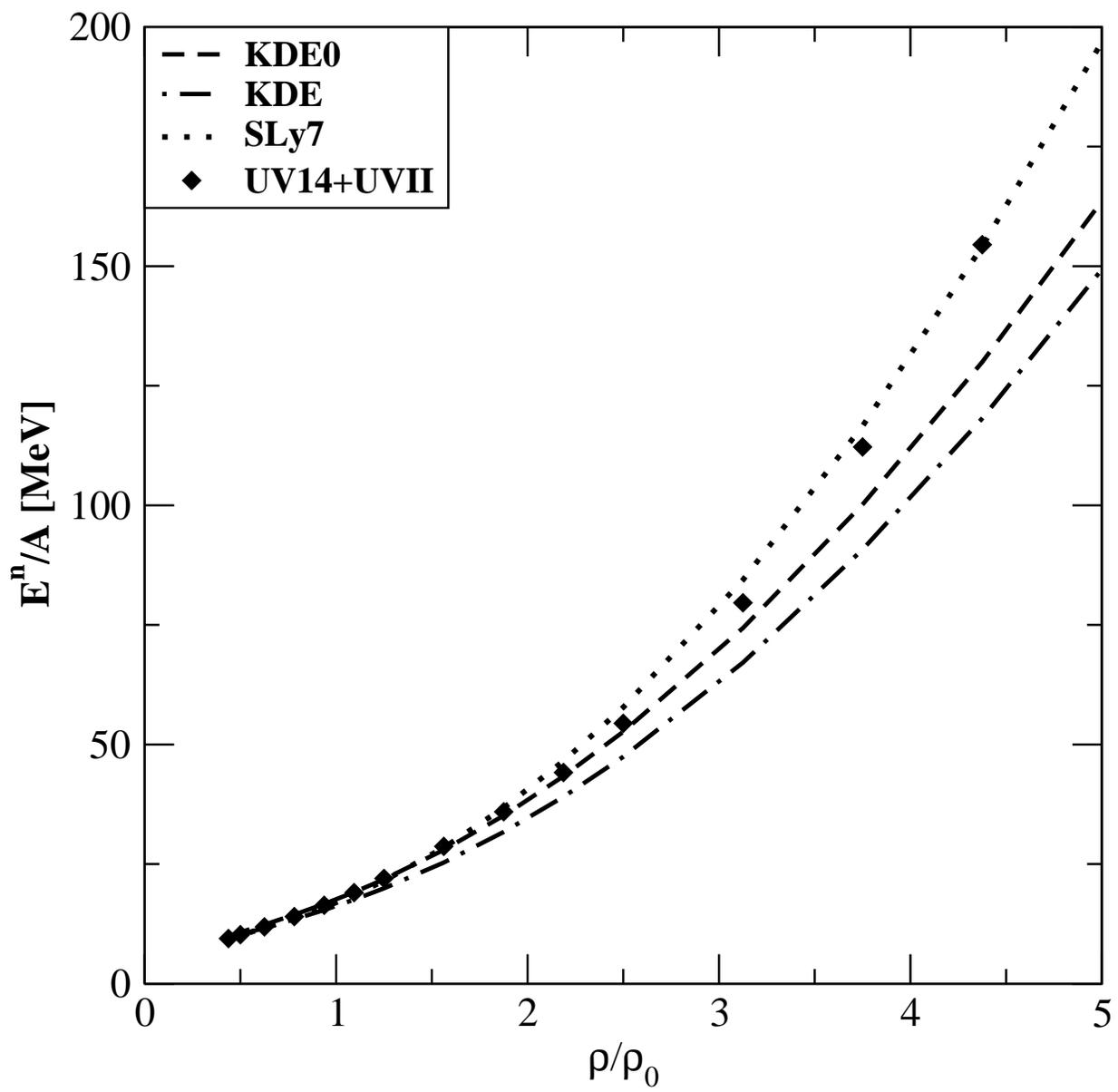
\end{document}